\documentstyle[aps,epsfig]{revtex} 
\begin{document}

{\Large\bf Kaon Production in Heavy-Ion 
Collisions and Kaon Condensation in Neutron Stars\footnote{Talks
presented at (1): Space-Time'97, East Lansing, USA, May 28-31, 1997,
(2): International Workshop on Hadrons in
Dense Matter, GSI, Darmstadt, July 2-6, 1997, and (3): 
International Conference on Physics since Parity Breaking, A Conference
in Memory of Professor C.S. Wu, Nanjing, China, August 16-18,1997.
To appear in the Proceedings of (3)}}\\[2mm]
{\large\em G.Q. Li, C.-H. Lee, and G.E. Brown}\\[2mm]
{\small Physics Department, State University of New York at
Stony Brook, Stony Brook, New York 11794, USA}\\[2mm]

The recent past witnesses the growing interdependence
between the physics of hadrons, the physics of relativistic
heavy-ion collisions, and the physics of compact objects in
astrophysics. A notable example is the kaon which plays special
roles in all the three fields. In this talk, we
first review the various theoretical investigations
of kaon properties in nuclear medium. We then present
a detailed transport model study of kaon production and flow in
heavy-ion collisions at SIS energies. 
Finally, We discuss the effects of the kaon in-medium 
properties extracted from heavy-ion data on neutron 
star properties, especially on the lowering of maximum 
mass of neutron stars with the onset of kaon condensation 
around $3\rho_0$.

\vskip 0.5cm

\section{INTRODUCTION}

There is currently growing interplay between physics of
hadrons (especially the properties of hadrons in 
dense matter which might reflect spontaneous
chiral symmetry breaking and its restoration), the physics
of relativistic heavy-ion collisions (from which one might 
extract hadron properties in dense matter), and the
physics of compact objects in astrophysics (which needs
as inputs the information gained from the first two fields).
A notable example is the kaon ($K$ and ${\bar K}$), which, 
being a Goldstone boson with strangeness, 
plays a special role in all the three fields mentioned.

Ever since the pioneering work of Kaplan and Nelson \cite{kap86}
on the possibility of kaon condensation in dense nuclear
matter, many works have been devoted to the study of kaon
properties in nuclear matter. There are two typical
approaches to this problem,
one based on the chiral perturbation theory 
\cite{brown87,wise91,lee94,lee95,kai96,lee96,kai97,waas97}, 
and the other based on the extension of the Walecka-type
mean field model from SU(2) to SU(3) \cite{sch94,knor95}. 
Although quantitatively,
the results from these different models are not
completely identical, qualitatively, a consistent 
picture, namely in nuclear matter the kaon feels a weak
repulsive potential and the antikaon feels a strong
attractive potential, has emerged. 

Measurements of kaon spectra and flow have been systematically
carried out in heavy-ion collisions at SIS (1-2 AGeV), 
AGS (10 AGeV), and SPS (200 AGeV) energies \cite{qm96}. 
Of special interest is kaon production in
heavy-ion collisions at SIS energies, as it has been shown
that particle production at subthreshold energies 
is sensitive to its properties in dense matter
\cite{cass90,koli96,kkl97}. 
 
Studies of neutron star properties also have a long history.
A recent compilation by Thorsett quoted by Brown \cite{nsmass}
shows that well-measured neutron star masses are all less
than 1.5$M_\odot$. On the other hand, most of the theoretical 
calculations based on conventional nuclear equation of state 
(EOS) predict a maximum neutron state mass above 2$M_\odot$. 
The EOS can, therefore, be substantially softened without running into 
contradiction with observation. Various scenarios have been 
proposed that can lead to a soft EOS. Brown and collaborators 
suggested that kaon condensation might happen at a critical 
density of 2-4$\rho_0$ \cite{brown94}. 

In the first part of this talk we review
various theoretical approaches for studying kaon 
properties in nuclear medium. In the second part we present
a detailed analysis of kaon production in
heavy-ion collisions at SIS energies. By comparing transport model
predictions with experimental data, we try to extract 
kaon in-medium properties at high densities. 
This information provides guidance
for the study of kaon condensation in neutron stars, which is
the focus of the third part of the talk. 

\section{kaon in dense matter: a review}

The interactions between pseudoscalar meson and baryon are usually 
described by the $SU(3)_L\times SU(3)_R$ nonlinear chiral
Lagrangian. In the mean-field approximation and including only the
Weinberg-Tomozawa and the Kaplan-Nelson terms,
the kaon and antikaon energies are given by
\begin{eqnarray}\label{omek}
\omega_{K,\bar K}=\left[m_K^2+{\bf k}^2-{\Sigma_{KN}\over f^2}\rho_S
+\left({3\over 8}{\rho_N\over f^2}\right)^2\right]^{1/2}
\pm {3\over 8}{\rho_N\over f^2}.
\end{eqnarray}
Note that the scalar attraction depends on nucleon scalar
density $\rho_S$, which is model dependent \cite{sch94}.
The scalar attraction is proportional to the kaon-nucleon sigma 
term, $\Sigma _{KN}$, which can be related to
pion-nucleon sigma term, $\Sigma _{\pi N}$. 
The latter is relatively well determined to be about 
45 MeV. There are, unfortunately, large uncertainties
in the strange quark mass and and nucleon strangeness content.
There are also a number of corrections to these simple 
mean-field results, such as the range terms which 
reduce the scalar attraction \cite{lee96}, 
the Brown-Rho scaling in $f_\pi$ \cite{br91,br96}, 
which cancels approximately the short-range
correlation effects \cite{pand95}, and the coupled channel effects
in the case of $K^-$ \cite{waas97,koch94}. A detailed discussion on
these can be found in Ref. \cite{lilee97}.

In view of large uncertainties in $\Sigma _{KN}$ and
difficulties in treating systematically high-order
corrections, we adopt here a more
phenomenological approach. We assume that the effects from 
the scaling in $f_\pi$ and short-range correlations
approximately cancel each other. Furthermore, we
introduce two free parameters, $a_K$ and $a_{\bar K}$, 
which determine the scalar attractions for $K^+$ and
$K^-$. The kaon and
antikaon energy in the nuclear medium can then be written as
\begin{eqnarray}
\omega _{K,\bar K}=\left[m_K^2+{\bf k}^2-a_{K,\bar K}\rho_S
+(b_K \rho_N )^2\right]^{1/2} + b_K \rho_N  ,
\end{eqnarray}
where $b_K=3/(8f_\pi^2)\approx 0.333$ GeVfm$^3$. 
We try to determine $a_K$ and $a_{\bar K}$ from the 
experimental observables in heavy-ion collisions.

\begin{figure}
\begin{minipage}[b]{0.46\linewidth}
\centering\epsfig{file=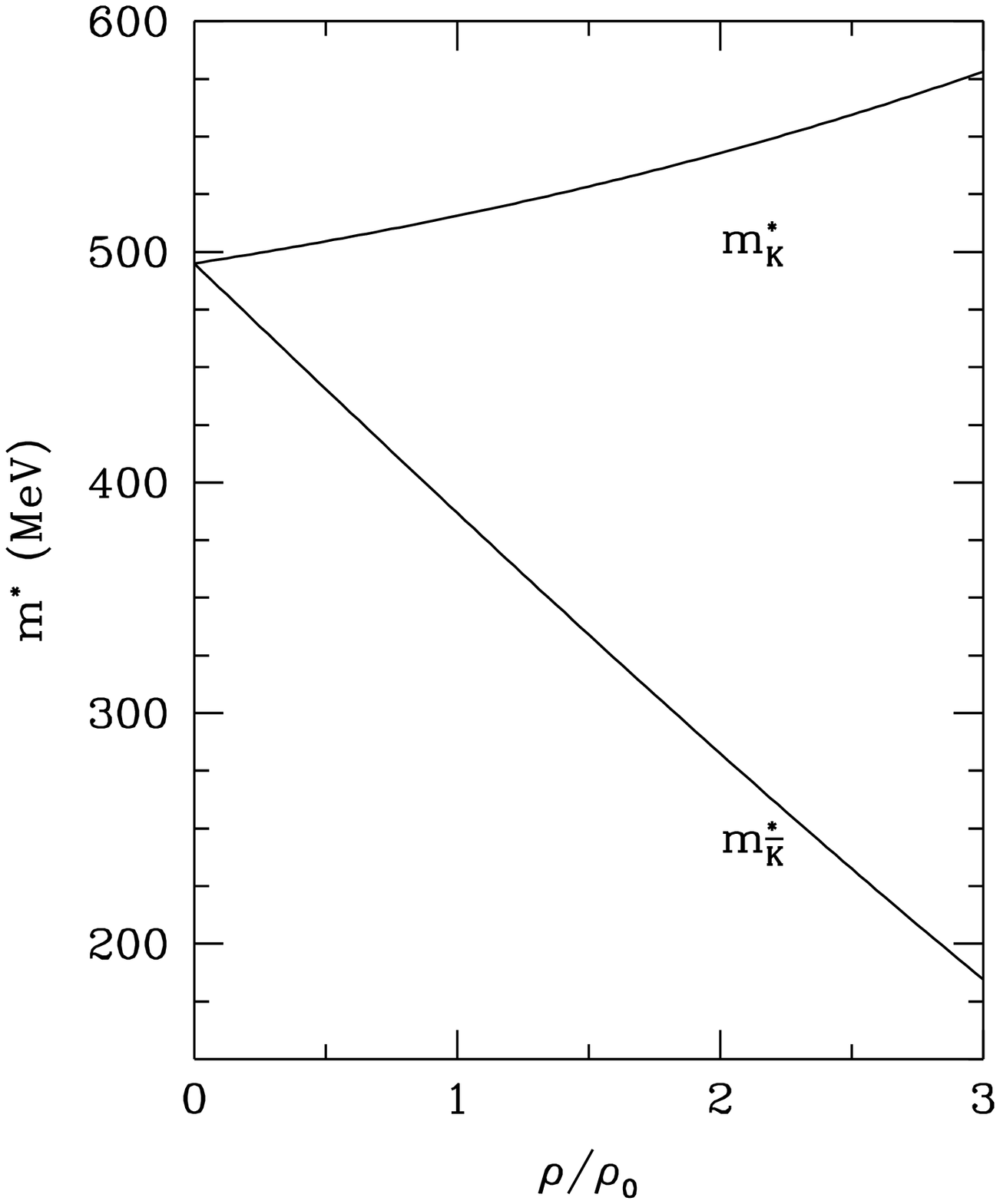,width=3.2in}
\caption{Effective masses of kaon and antikaon.}
\label{kmass}
\end{minipage}\hfill
\begin{minipage}[b]{0.46\linewidth}
\centering\epsfig{file=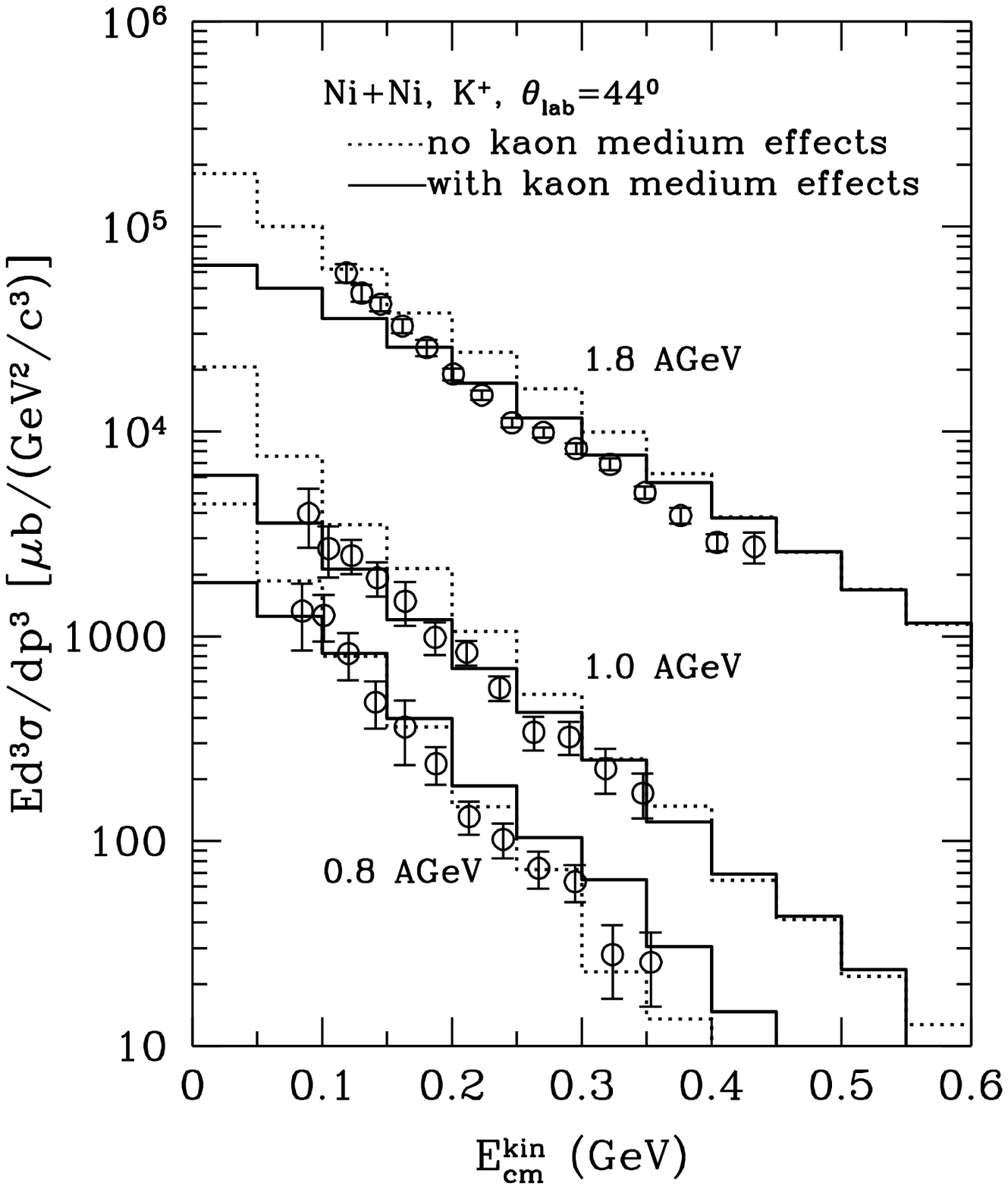,width=3.2in}
\caption{$K^+$ kinetic energy spectra.}
\label{kkinall}
\end{minipage}
\end{figure}

Since the kaon-nucleon interaction is relatively weak as
compared to other hadron-nucleon interactions,
impulse approximation is considered to be reasonable
for the kaon potential in nuclear matter, at least at 
low densities. This provides some constraint on $a_K$.
Using $a_K=0.22$ GeV$^2$fm$^3$, we find that at normal
nuclear matter density $\rho_0 =0.16$ fm$^{-3}$,
the $K^+$ feels a repulsive potential of about 20 MeV.
This is in rough agreement with what is expected from
the impulse approximation using the $KN$ scattering length
in free space. Determination of the $a_{\bar K}$ is more delicate, as
impulse approximation does not apply. 
We find that $a_{\bar K}=0.45$ GeV$^2$fm$^3$
provides a good fit to the $K^-$ data in heavy-ion
collisions at SIS energies. 
With these two parameters we show in Fig. \ref{kmass} the effective 
masses of kaon and antikaon defined as their energies at zero momentum. 
It is seen that the kaon mass increases slightly with density, 
while that of the antikaon drops substantially. 

\section{kaon production in heavy-ion collisions}

One of the most important ingredients in the transport
model study of particle production in heavy-ion
collisions is the elementary particle production
cross sections in hadron-hadron interactions. At SIS
energies, the colliding system consists mainly of
nucleons, delta resonances, and pions. We need thus
kaon and antikaon production cross sections from nucleon-nucleon
($NN$), nucleon-delta ($N\Delta$), delta-delta ($\Delta\Delta$),
pion-nucleon ($\pi N$), and pion-delta ($\pi\Delta$)
interactions. In addition, the antikaon can also be produced
from strangeness-exchange processes such as $\pi Y\rightarrow 
{\bar K} N$. Because of the lack of experimental data,
especially near production threshold that are important
for heavy-ion collisions at SIS energies, we have to adopt
the strategy that combines the parameterization of experimental
data with some theoretical investigations and reasonable assumptions
and prescriptions. A comprehensive analysis of these
elementary cross section can be found in Ref. \cite{lilee97}.
 
Particles produced in elementary hadron-hadron 
interactions in heavy-ion collisions cannot 
escape the environment freely and be detected. Instead, they
are subjected to strong final-state interactions.
For the kaon, because of strangeness conservation,
its scattering with nucleons at low energies is
dominated by elastic and pion production processes,
which do not affect its final yield but changes its momentum 
spectra. The final-state interaction for the antikaon is much 
stronger. As mentioned, antikaons can be destroyed in
the strangeness-exchange processes. They also undergo 
elastic scattering. 

\begin{figure}
\begin{minipage}[b]{0.46\linewidth}
\centering\epsfig{file=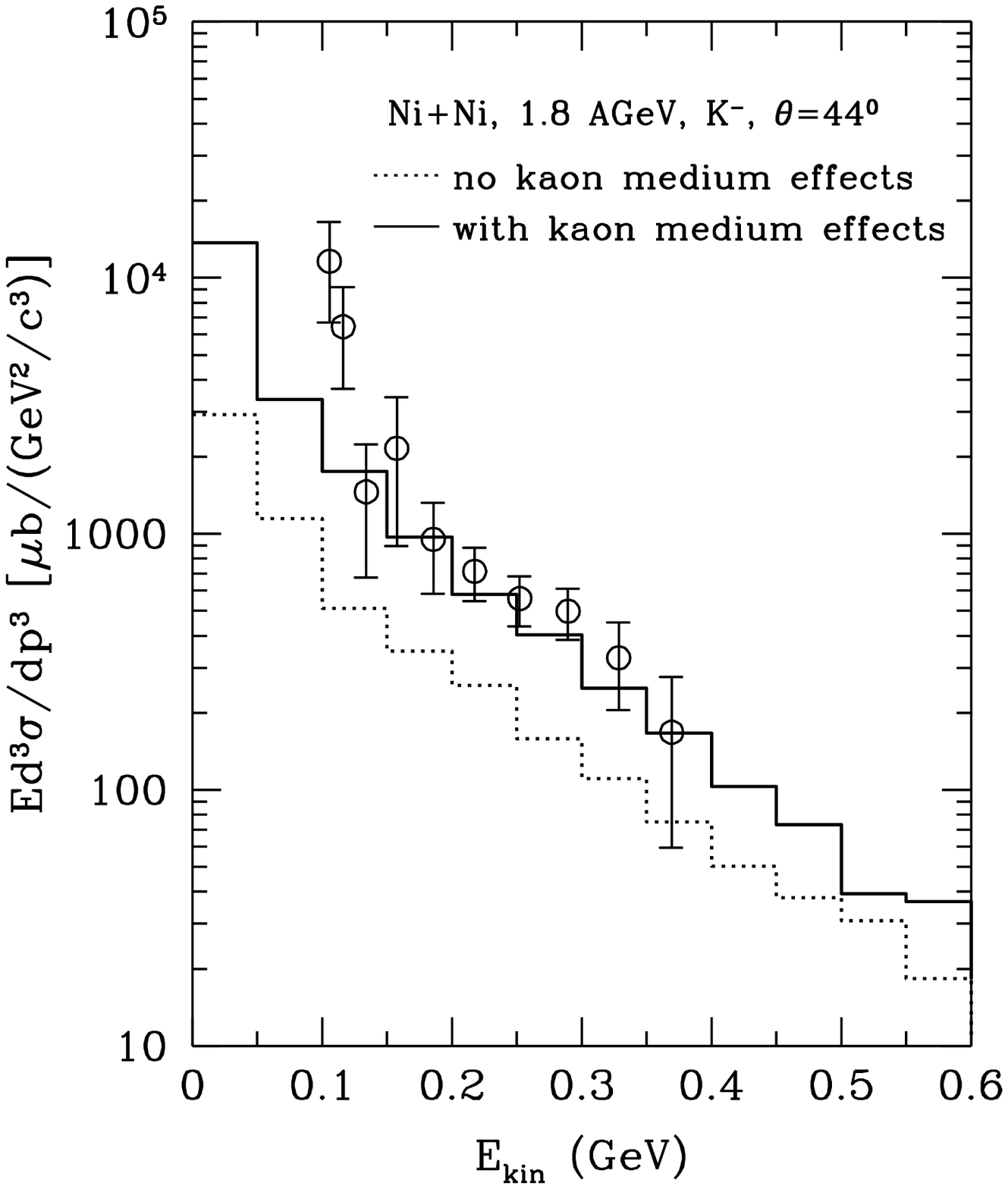,width=3.2in}
\caption{$K^-$ kinetic energy spectra.}
\label{akni18}
\end{minipage}\hfill
\begin{minipage}[b]{0.46\linewidth}
\centering\epsfig{file=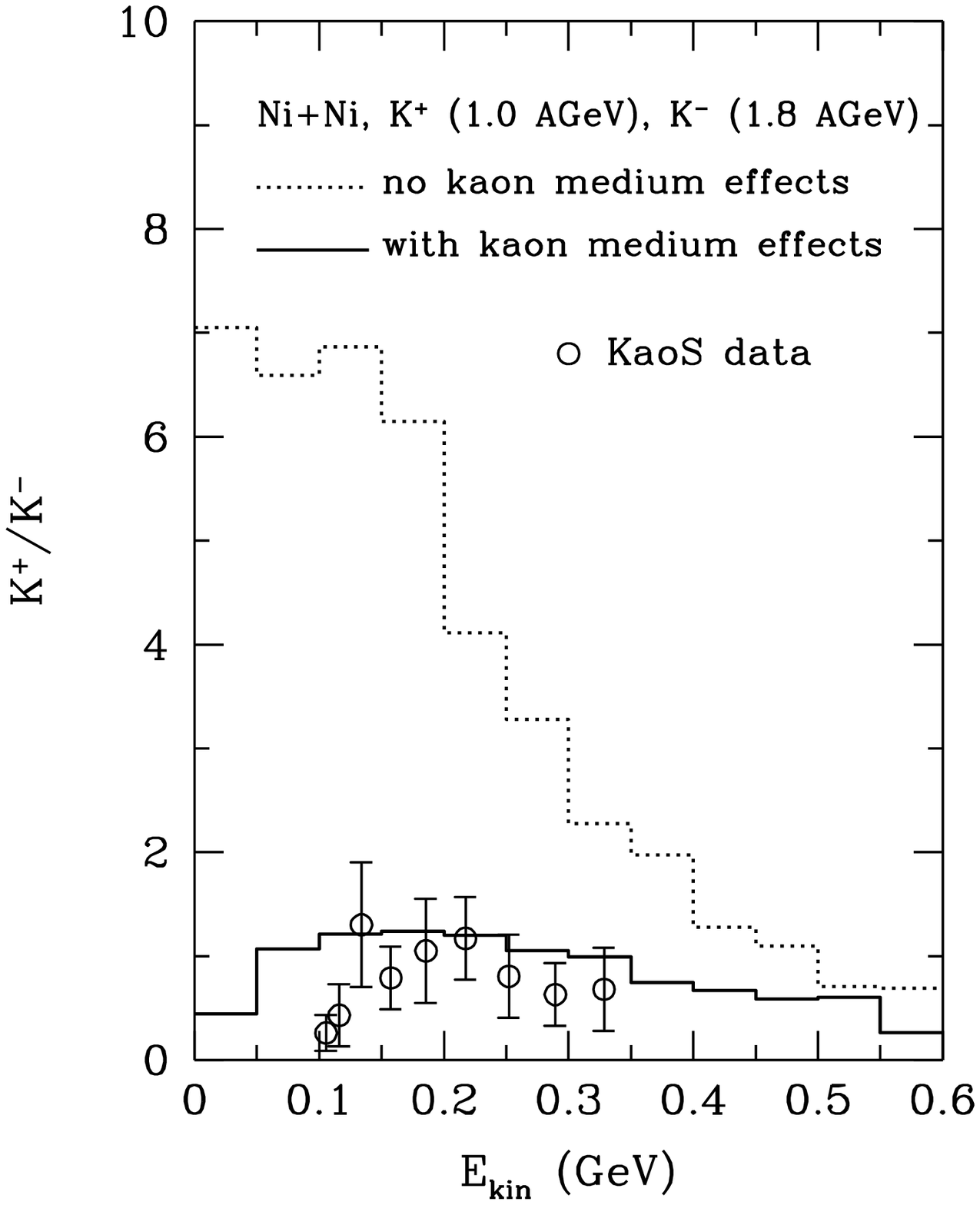,width=3.2in}
\caption{$K^+/K^-$ ratio as a function of kinetic energy.}
\label{kakr}
\end{minipage}
\end{figure}

The results for $K^+$ kinetic energy 
spectra in Ni+Ni collisions at 0.8, 1.0 and 1.8 AGeV are
shown in Fig. \ref{kkinall}. The solid histogram gives
the results with kaon medium effects, while the dotted 
histogram is the results without kaon medium effects. 
The open circles are the experimental data from the KaoS 
collaboration {\cite{kaos}}. It is seen that 
the results with kaon medium effects are in good agreement 
with the data, while those without kaon medium effects slightly 
overestimate the data. We note that kaon feels a small 
repulsive potential; thus the inclusion of the kaon medium 
effects reduces the kaon yield. The slopes of the kaon spectra 
in the two cases also differ. With a repulsive potential, 
kaons are accelerated during the propagation, leading to a 
larger slope parameter as compared to the case without
kaon medium effects.

The results for the $K^-$ kinetic energy spectra are shown 
in the middel window Fig. \ref{akni18} for Ni+Ni collision at 1.8 AGeV. 
It is seen that without medium effects, our
results are about a factor 3-4 below the experimental data.
With the inclusion of the medium effects which reduces the
antikaon production threshold, the $K^-$ yield increases by about a factor
of 3 and our results are in good agreement with the data. This
is similar to the findings of Cassing {\it et al.} \cite{cass97a}.
Also, with the inclusion of kaon medium effects, the slope parameter
of the $K^-$ spectra decreases, since the propagation of
antikaons in their attractive potential reduces their
momenta.

The medium effects on kaon and antikaon can be best seen
by looking at their ratio as a function
of kinetic energy. This is shown
Fig. \ref{kakr}. Without kaon medium effects, the $K^+/K^-$
ratio decreases from about 7 at low kinetic energies to
about 1 at high kinetic energies. 
Since the antikaon absorption cross section by nucleons
increases rapidly at low momentum, low-momentum
antikaons are more strongly absorbed by nucleons than 
high-momentum ones. This makes the $K^+/K^-$ ratio increase
with decreasing kinetic energies. When medium effects are
included, we find that the $K^+/K^-$ ratio is almost
a constant of about 1 in the entire kinetic energy region,
which is in good agreement with the experimental data from
the KaoS collaboration \cite{kaos}, shown in the figure
by open circles. 

\begin{figure}
\begin{minipage}[b]{0.46\linewidth}
\centering\epsfig{file=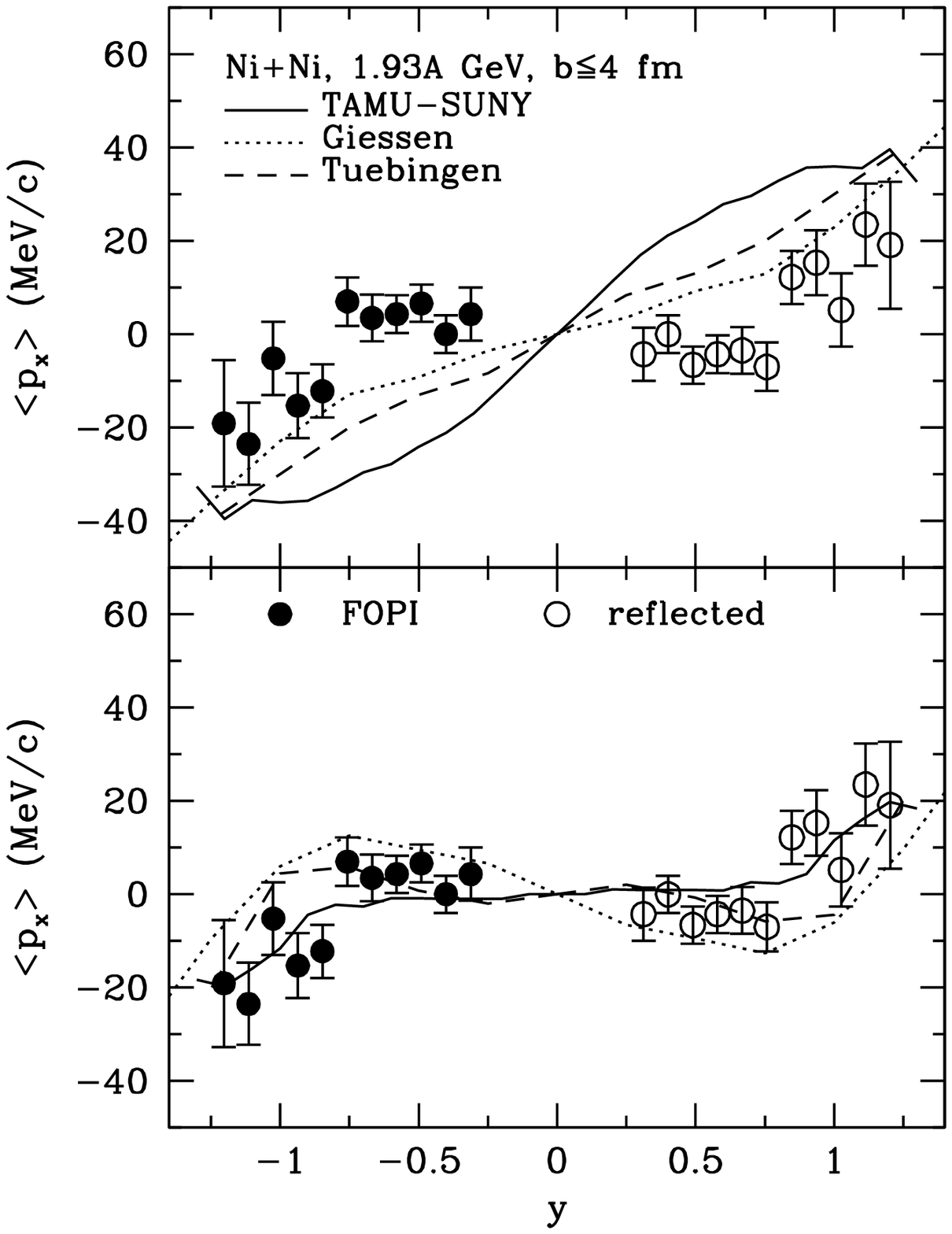,width=3.2in}
\caption{$K^+$ flow from different calculations}
\label{kflow}
\end{minipage}\hfill
\begin{minipage}[b]{0.46\linewidth}
\centering\epsfig{file=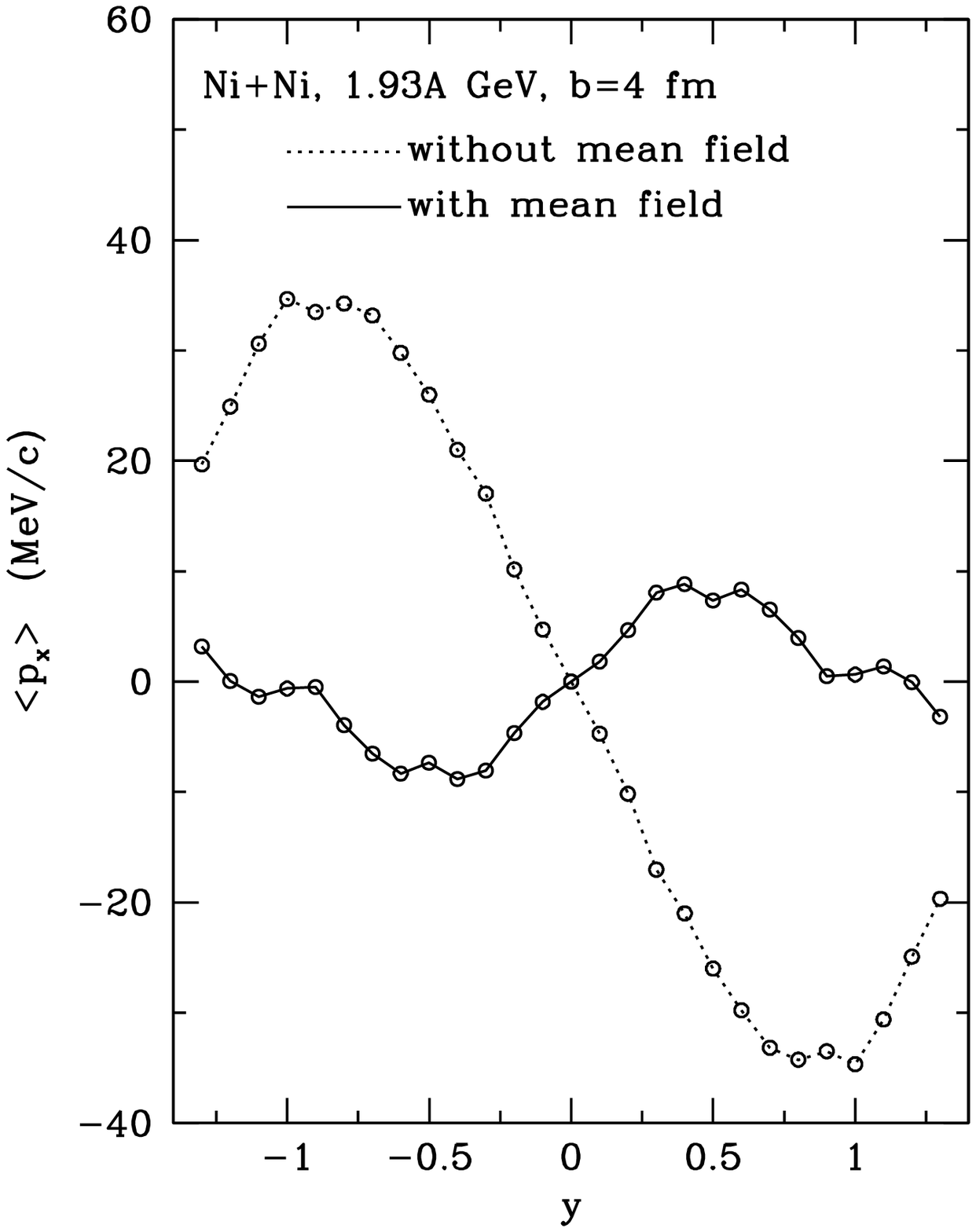,width=3.2in}
\caption{$K^-$ flow in Ni+Ni collisions.}
\label{akflow}
\end{minipage}
\end{figure}

Another observable that can probe kaon potential in
dense matter more clearly is kaon flow, which is
not affected by the uncertainties in the elementary 
cross sections \cite{likoli95}. The results for $K^+$ flow in Ni+Ni
collisions from different transport models are
compared with experimental data \cite{ritman}
in Figs. \ref{kflow}. The results shown in the left window
are obtained without kaon potential, while those in the right window
include the kaon potential. The solid, dashed,
and dotted curves are from Refs. \cite{likoli95}, \cite{cass97b}, and
\cite{fae97}, respectively. It is seen that, within the experimental
errorbars, the results with a weak repulsive kaon potential is
in better agreement with the data. 

Similarly one can analyse $K^-$ flow. The results from
Ref. \cite{liko96} are shown in Fig. \ref{akflow}. The solid
and dashed curves are obtained with and without $K^-$
potential, respectively. Without kaon potential, we see
that antikaons flow in the opposite direction to that of nucleon,
simply because of the strong absorption of antikaons
by spectator nucleons. Once the attractive $K^-$ potential is
turned on, those surviving antikaons are pulled towards
nucleons, resulting in a weak flow signal. Primary experimental
data seem to be in better agreement with the scenario 
with antikaon potential.
 
\section{kaon condensation in neutron stars}

\begin{figure}
\begin{minipage}[b]{0.46\linewidth}
\centering\epsfig{file=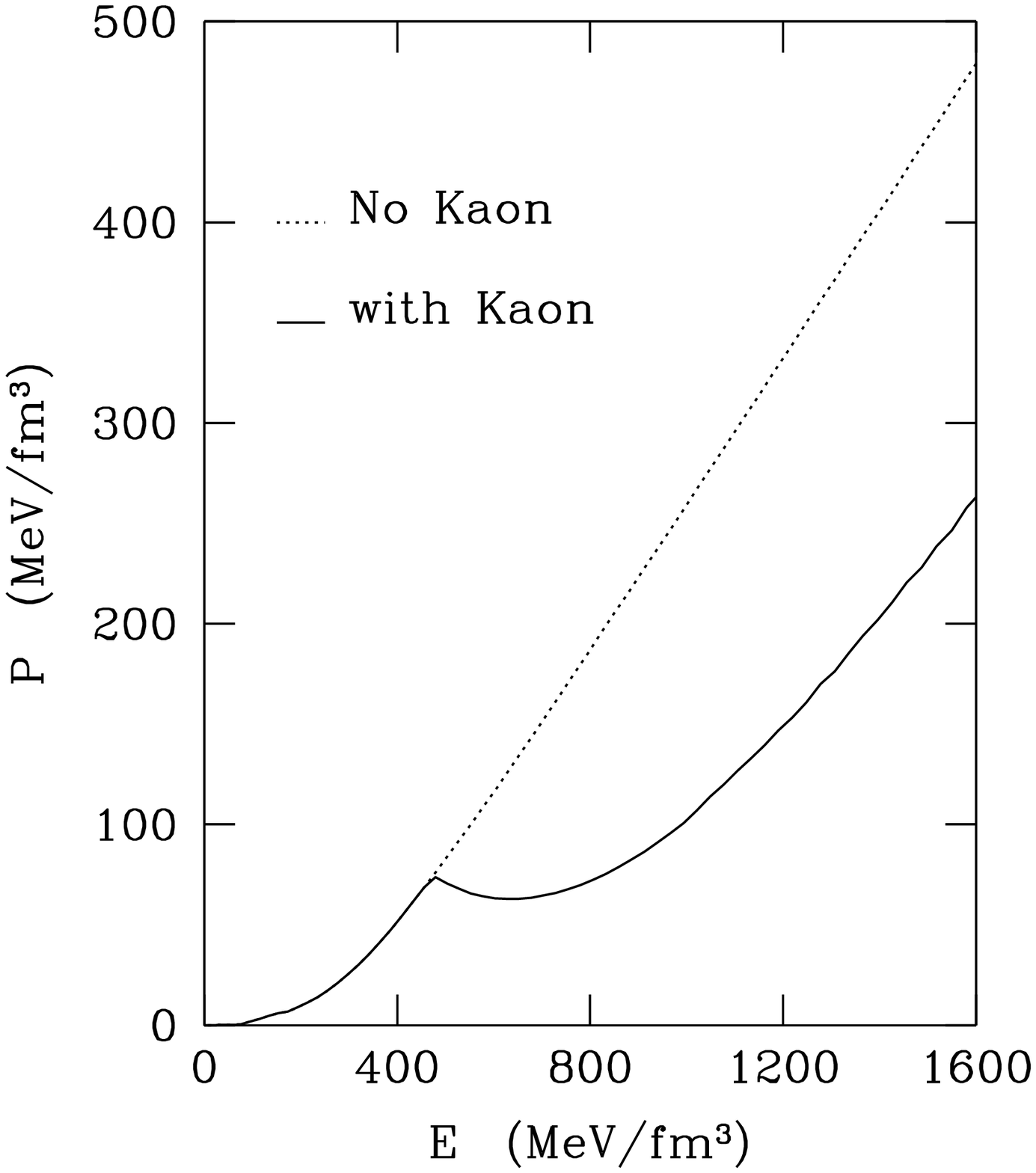,width=3.2in}
\caption{Equation of state of neutron star matter.}
\label{pres}
\end{minipage}\hfill
\begin{minipage}[b]{0.46\linewidth}
\centering\epsfig{file=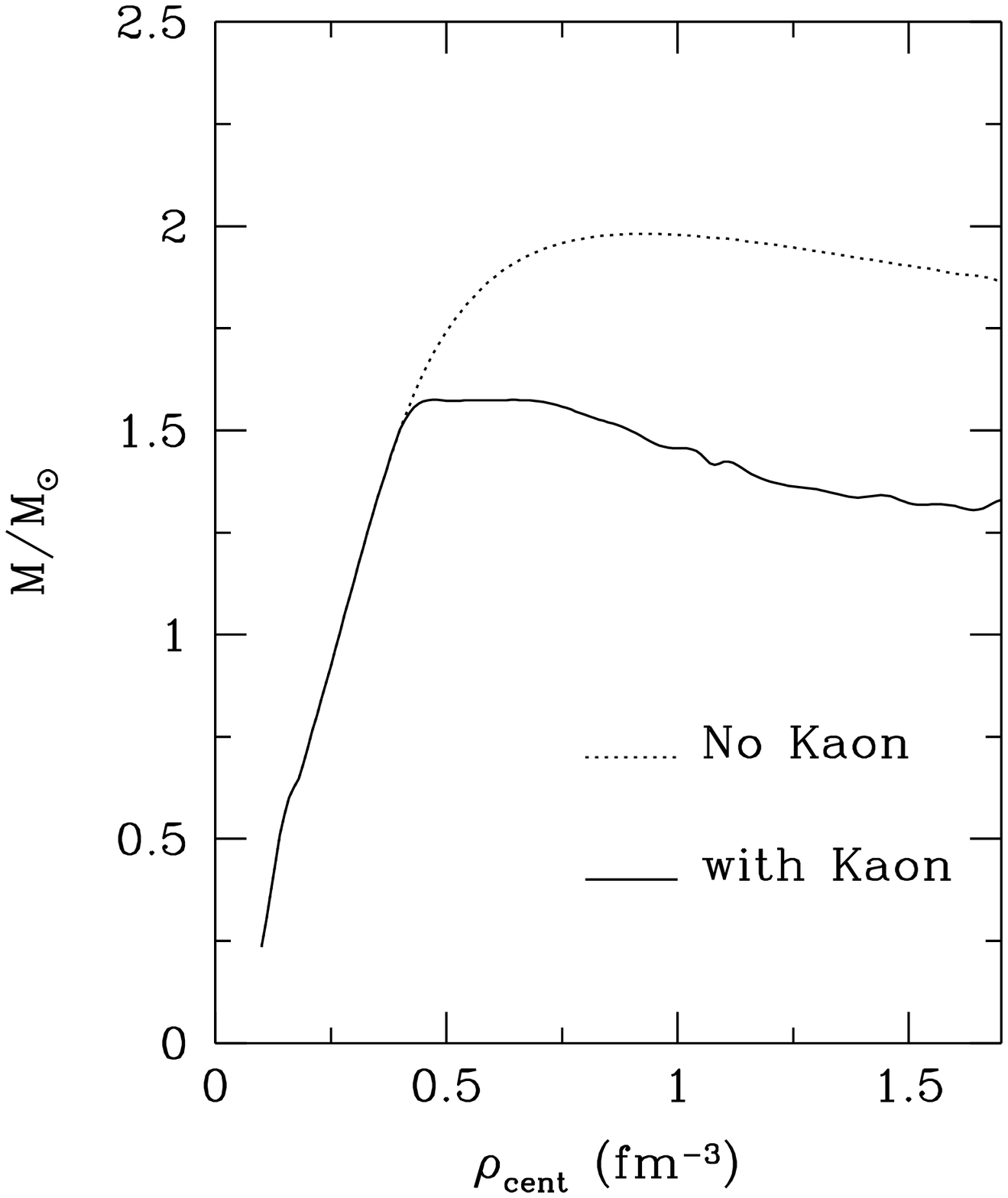,width=3.2in}
\caption{Neutron star mass as a function of central density}
\label{mass}
\end{minipage}
\end{figure}

As mentioned, the medium modification of kaon properties
affects not only kaon observables in heavy-ion collisions,
it also bears important consequences in the structure and
evolution of compact objects in astrophysics, especially
the maximum mass of neutron stars. 
In the presence of kaons, the chemical equilibrium conditions
require that the chemical potentials should satisfy
\begin{eqnarray}
\mu = \mu_n-\mu_p = \mu _e = \mu_\mu =\mu _{K^-}.
\end{eqnarray}
The local charge neutrality can be imposed by minimizing
the thermodynamical potential
\begin{eqnarray}
\Omega = \varepsilon _N + \varepsilon _{K^-}
+\varepsilon _L - \mu (\rho _p -\rho _{K^-} - \rho _e -\rho _\mu ).
\end{eqnarray} 
For the energy density of nucleons we use the results
of Furnstahl, Tang, and Serot \cite{fst}. 
The energy density $\varepsilon$ of the ground state of the 
system is obtained by extremizing $\Omega$.
The pressure of the system is then obtained from the energy
density. They are then used in the TOV equation to obtain the properties
of neutron stars. The pressure as a function of energy density
for both the cases with and without kaons are shown in
Fig. \ref{pres}.

The results for neutron star mass as a function of central
density $\rho _{cent} $ are shown in Fig. \ref{mass}. It is seen
that, without kaons, the maximum neutron star mass in 
this model is about 2$M_\odot$. Similar 
conclusions, with neutron star mass in the range of 
2.1-2.3$M_\odot$, have been obtained in Ref. 
\cite{eng96} based on the nuclear equation of state
from the Dirac-Brueckner-Hartree-Fock approach. 
When kaons are included, the maximum mass of neutron 
stars is about 1.5$M_\odot$. 

\section{summary and outlook}

In summary, we studied $K^+$ and $K^-$ production 
in Ni+Ni collisions at 1-2 AGeV, based on the relativistic 
transport model including the strangeness degrees of freedom. 
We found that the recent experimental data from the KaoS
collaboration are consistent with the predictions of the chiral
perturbation theory that the $K^+$ feels a weak repulsive potential
and $K^-$ feels a strong attractive potential in nuclear 
medium. Using the kaon in-medium properties constrained
by the heavy-ion data, we have studied neutron star properties
with and without kaon condensation. The maximum mass of 
neutron stars is found to be about 2.0$M_\odot$ based 
on conventional nuclear equations of state 
obtained from the effective Lagrangian of Furnstahl {\it et al.}. 
This can be reduced to about 1.5$M_\odot$,
once kaon condensation is introduced.
We have emphasized the growing interdependence
between hadron physics, relativistic heavy-ion physics
and the physics of compact stars in astrophysics.

We are grateful to C.M. Ko and M. Rho 
for helpful discussions, and N. Herrmann and P. Senger 
for useful communications. This work is supported in part 
by the Department of Energy under Grant No. DE-FG02-88ER40388.

\end{document}